\documentclass[twocolumn,preprintnumbers,amsmath,amssymb,superscriptaddress,footnote]{revtex4-2}
\usepackage{graphicx} 
\usepackage[dvipdfmx]{epsfig} 
\usepackage{bm}
\usepackage{ulem}
\usepackage{amsmath,color}

\def\m@thcombine#1#2{
  \setbox0=\hbox{$#1$}
  \setbox1=\hbox{$#2$}
  \ifdim\wd0>\wd1
    \setbox0=\hbox to\wd1{\hss\box0\hss}
  \else
    \setbox1=\hbox to\wd0{\hss\box1\hss}
  \fi
  \mathop{\vcenter{
    \offinterlineskip\box0\box1}}}
\def\lesim{\m@thcombine<\sim}
\def\gesim{\m@thcombine>\sim}

\begin{document}
\title{Imprints of $\alpha$ clustering in the density profiles of $^{12}$C and $^{16}$O}
\author{W. Horiuchi}
\email{whoriuchi@omu.ac.jp}
\affiliation{Department of Physics, Osaka Metropolitan University, Osaka 558-8585, Japan}
\affiliation{Nambu Yoichiro Institute of Theoretical and Experimental Physics (NITEP), Osaka Metropolitan University, Osaka 558-8585, Japan}
\affiliation{RIKEN Nishina Center, Wako 351-0198, Japan}
\affiliation{Department of Physics,
  Hokkaido University, Sapporo 060-0810, Japan}

\author{N. Itagaki}
\email{itagaki@omu.ac.jp}
\affiliation{Department of Physics, Osaka Metropolitan University, Osaka 558-8585, Japan}
\affiliation{Nambu Yoichiro Institute of Theoretical and Experimental Physics (NITEP), Osaka Metropolitan University, Osaka 558-8585, Japan}

\begin{abstract}
  The $^{4}$He nucleus is a well bound and highly correlated
    four-nucleon system that is also found in form of $\alpha$-clusters
    as substructure in nuclear many-body systems. The standard
    single-particle shell model cannot represent these four-body correlations.
    It is highly desirable to find a simple way to identify and
    distinguish these fundamental structures without large-scale computations.
    In this paper, we investigate with intrinsic Slater determinants as trial
    states in how far these two competing pictures prevail in the ground
    states of $^{12}$C and
    $^{16}$O. The trial states
  can describe both the $j$-$j$ coupling shell-model
  and $\alpha$-cluster configurations in a unified way.
  One-body density distributions of the trial states
    are calculated and compared to elastic scattering cross section data.
    The parameters of the trial state are chosen to reproduce the experimental
    charge radii.
   A well developed cluster structure is characterized by an enhancement
  of the differential elastic scattering
  cross sections, as well as the elastic charge form factors
  around the first peak positions.
  The density profiles of the internal regions can also be probed
  at higher momentum transfer
  regions beyond the second minimum.
  With these trial states one can visualize the competition between cluster and shell structure in an intuitive and simple way.
\end{abstract}

\maketitle

How can we distinguish shell and cluster states quantitatively?
This is a fundamental question of nuclear structure physics.
Each nucleus is composed of protons and neutrons,
which in the single-particle shell model,
are considered to perform independent particle motions
in a self-consistent potential~\cite{Mayer55}.
On the other hand, two protons and two neutrons 
often form an $\alpha$ cluster as the binding energy of $^4$He is so large
compared to the neighboring nuclei.
These two different aspects,
the cluster and shell picture,
compete in the nuclear structure~\cite{Brink}.

From the point of view of the nucleon-nucleon interaction, the non-central interactions,
the spin-orbit and tensor interactions, play significant roles in the nuclear systems, which are the key ingredients for this competition. The spin-orbit interaction is a driving force that stabilizes the shell-model picture
(the magic number can be explained owing to this effect) and break $\alpha$ clusters~\cite{Itagaki04},
whereas the strong binding of $^4$He comes from the tensor contribution
which stabilizes the $\alpha$-cluster structure~\cite{ATMS}.
The balance of these non-central interactions governs the competition of the two pictures~\cite{Ishizuka22}.

For $^{12}$C and $^{16}$O, there have been numerous works based on
the three- and four-$\alpha$ cluster
models, respectively~\cite{Uegaki77,Kamimura81,Descouvemont93,Itagaki95,THSR}.  
However, from the shell-model point of view,
these nuclei correspond to the closures of 
$p_{3/2}$ and $p_{1/2}$ subshells, respectively.
In $^{12}$C, the attractive effect of the spin-orbit interaction
breaks the $\alpha$ clusters according to the shell-model picture,
but the $\alpha$ cluster model succeeded in 
explaining many electromagnetic properties.
An {\it ab initio} computation was carried out to solve
this 12-nucleon problem and succeeded in reproducing
the elastic charge form factor of $^{12}$C~\cite{Lovato13}.
The recent {\it ab initio} shell-model calculation showed
the mixing of the shell and cluster configurations~\cite{Otsuka22}, 
as suggested by molecular dynamics calculations~\cite{Kanadaenyo07,Chernykh07}.
For $^{16}$O, if we take a small distance between 
four $\alpha$ clusters with a tetrahedron configuration,
the wave function agrees with the shell model.
In contrast, recent {\it ab initio} calculations suggested
the possibility of four-$\alpha$ clusters with
finite relative distance for the ground state~\cite{Epelbaum14}.

It is highly desirable to have a simple picture
  in order to estimate if $\alpha$-cluster or shell-model structure
  dominates nuclear states
as it impacts
the astrophysically important reaction rates, e.g., triple-$\alpha$ reactions
involving the so-called Hoyle state~\cite{Salpeter52,Hoyle54}.
For this purpose, the basis states of the antisymmetrized
 quasi-cluster model (AQCM)~\cite{AQCM01,AQCM02,AQCM03,AQCM04,AQCM05,AQCM06,AQCM07,AQCM08,AQCM09,AQCM10,AQCM11,AQCM12,AQCM13,AQCM14}
 are quite helpful as  they provide an ansatz for a simple trial state that is capable of representing both shell-model and cluster states
  on the same footing.

  In the following, we choose the parameters of these states 
  such that they represent either a pure $\alpha$ cluster, a single-particle
  shell-model state, or something in between. As there is no Hamiltonian
  or eigenvalue problem involved, nor a Ritz variational method
  to find optimal parameters, we require instead, that each trial state
  reproduces the experimental charge radius. This is possible by adjusting
  the spatial parameters of the two states as explained in detail below.
  In this way, any ambiguities coming from a model Hamiltonian or the choice
  of a many-body Hilbert space are absent.

  The one-body density calculated with a thus constrained trial state is
  used to calculate the proton-nucleus elastic scattering cross section
  and the elastic charge form factor which are then compared to measured data.
  This procedure allows to judge which trial state gives a better description
  of the data and hence which picture is closer to nature.

The shell and cluster trial states are represented
based on the $N\alpha$ cluster model~\cite{Brink}
and its extension: the antisymmetrized quasi-cluster model (AQCM~\cite{AQCM01,AQCM02,AQCM03,AQCM04,AQCM05,AQCM06,AQCM07,AQCM08,AQCM09,AQCM10,AQCM11,AQCM12,AQCM13,AQCM14}).
The trial state
is fully antisymmetrized ($\mathcal{A}$) and
is expressed by the product of the $\alpha$ particle
wave functions, $\Phi_{\alpha}$, as
\begin{align}
  \Phi(\nu,d,\Lambda)=\mathcal{A}
  \left\{\prod_{i=1}^N \Phi_\alpha(\nu,R_i,\Lambda)\right\},
\end{align}
Note that the inter-$\alpha$-cluster distance is
defined by $d=|\bm{R}_i-\bm{R}_j|$ ($i\neq j$),
where the three $\alpha$'s ($^{12}$C)
and four $\alpha$'s ($^{16}$O)
are placed with equilateral triangular and tetrahedron shapes,
respectively.
The wave function of the $i$th $\alpha$ particle 
with the Gaussian center parameter
$\bm{R}_i$, is defined by the product of the single-particle Gaussian wave packet as
\begin{align}
  \Phi_{\alpha}(\nu,R_i,\Lambda)=
  \phi^\nu_1(\uparrow,p)\phi^\nu_2(\downarrow,p)\phi^\nu_3(\uparrow,n)\phi^\nu_4(\downarrow,n)
\end{align}
with a single-nucleon Gaussian wave packet
with spin $\chi_s$ ($s=\uparrow$ or $\downarrow$)
and isospin $\eta_t$ ($t=p$ or $n$) wave functions
\begin{align}
  \phi^\nu_j(s,t)=  \left(\frac{2\nu}{\pi}\right)^{3/4}\exp\left[-\nu(\bm{r}_j-\bm{\zeta}_i)^2\right]\chi_{s}\eta_{t},
\end{align}
where
\begin{align}
  \bm{\zeta}_i=\bm{R}_i+i\Lambda \bm{e}^{\rm spin}\times \bm{R}_i
\end{align}
with $\bm{e}^{\rm spin}$ being a unit vector for the intrinsic-spin
orientation of a nucleon.
Apparently, $\Lambda=0$ leads to 
the ordinary Brink-type $\alpha$-cluster basis function~\cite{Brink}.
This multi-$\alpha$-cluster basis function can also describe the
shell-model configuration.
A limit of $d\to 0$ leads to the SU(3) limit of the shell-model configuration,
$(0s)^4(0p)^8$ and $(0s)^4(0p)^{12}$ for $^{12}$C and $^{16}$O, respectively.
The $\Lambda$ value controls the degree of the breaking of the $\alpha$ particles.
As was shown in Ref.~\cite{AQCM04},
the $j$-$j$ coupling shell-model closure configuration for $^{12}$C,
$(0s_{1/2})^4(0p_{3/2})^8$, can be expressed by taking $\Lambda=1$ with $d\to 0$.

Once the parameters of the trial function, i.e., $\nu$, $d$, and $\Lambda$,
are set, it is straightforward to evaluate the one-body
density distribution by
\begin{align}
  \tilde{\rho}(\bm{r})
  = \langle \Phi |\sum_{i=1}^A \delta(\bm{r}_i-\bm{r}) | \Phi \rangle / \langle \Phi | \Phi \rangle,
\end{align}
where $A$ is the  mass number.
Note $\sum_{i=1}^A\left<\bm{r}_i\right>=0$.
As the wave function is expressed by a Slater determinant,
this density distribution, in general, includes
the center-of-mass motion, which crucially affects the density profiles
in such light nuclei.
Since the present wave function can be deduced to
the product of the intrinsic and center-of-mass parts,
the center-of-mass wave function can exactly be factored out by
using a Fourier transform as~\cite{Horiuchi07}
\begin{align}
  \int d\bm{r}\,e^{i\bm{k}\cdot\bm{r}}\rho_{\rm int}(\bm{r})=\exp\left(\frac{k^2}{8A\nu}\right)
  \int d\bm{r}\,e^{i\bm{k}\cdot\bm{r}} \tilde{\rho}(\bm{r}).
\end{align}
The density distribution in the laboratory frame
is finally obtained by averaging that center-of-mass-free
intrinsic density distribution over angles as~\cite{Horiuchi12}
\begin{align}
  \rho(r)=\frac{1}{4\pi}\,\int d\hat{\bm{r}}\,\rho_{\rm int}(\bm{r}).
\label{density.eq}
\end{align}

The proton-nucleus elastic scattering cross sections are evaluated
by using the Glauber model~\cite{Glauber}. With the help of
the optical limit approximation~\cite{Glauber,Suzuki03},
the inputs to the theory are the density distribution obtained by Eq.~(\ref{density.eq})
and the profile function, which describes the nucleon-nucleon scattering properties~\cite{Ibrahim08}.
The elastic Coulomb phase is included in the calculation. 
The validity of the model is well tested, which can be seen in Refs.~\cite{Horiuchi16, Hatakeyama19}.
For more details, see, for example, Ref.~\cite{Horiuchi22c} and references therein,
showing the most recent application of this model.

\begin{table}[htb]
  \begin{center}
    \caption{Properties of the shell-model (S-type)
      and $\alpha$-cluster-model (C-type) trial states.
      Values in parentheses are obtained with the ideal shell-model configurations.
      The root-mean-square radii of $^{12}$C and $^{16}$O are adjusted to reproduce the experimental values 2.33 and 2.57 fm~\cite{Angeli13}, respectively}.
\begin{tabular}{lccccccc}
\hline\hline
  &&$\nu$ (fm$^{-2}$) &$d$ (fm)&$\Lambda$&$\left<Q\right>$ &$\left<LS\right>$&$\left<P\right>$\\
\hline
$^{12}$C &S-type&0.1886&0.010&1&8.0(8)&4.0(4)&$-4.0$($-4$) \\
        &C-type&0.2656&2.636&0&10.6 & 0.0(0) &$-2.0$ \\
        &M-type&0.2222 &2.000&0.2 &9.0 & 1.8  &$-2.9$ \\
\hline
$^{16}$O &S-type&0.1635&0.010&1&12.0(12)&0.0(0)& $-8.0$($-8$) \\
        &C-type&0.2656&3.058&0&18.6&$0.0$(0) &$-2.7$ \\
\hline\hline
\end{tabular}  
\label{results.tab}
\end{center}
\end{table}

To obtain the shell and cluster density profiles,
we extend the prescription given in Ref.~\cite{Horiuchi22c}
and adapt it to the present multi-$\alpha$ cases.
For the shell-model density profile (S-type),
we take a small $\alpha$-cluster distance $d=0.01$ fm
and $\Lambda=1$,
which gives shell-model configurations of $(0s_{1/2})^4(0p_{3/2})^8$ for $^{12}$C
and $(0s_{1/2})^4(0p_{3/2})^8(0p_{1/2})^4$ for $^{16}$O.
Note that in the $^{16}$O case the result does not depend on
  the choice of $\Lambda$ as the $p$-shell is completely filled.
The remaining parameter $\nu$ is fixed
so as to reproduce the charge radius~\cite{Angeli13}.

For the $\alpha$-cluster density profiles (C-type),
the $\nu$ value is taken commonly to reproduce
the charge radius of $\alpha$ particle, $\nu=0.2656$~fm$^{-2}$,
and we fix the $d$ value to reproduce the charge radii of $^{12}$C or $^{16}$O,
where they are assumed to have equilateral triangular
and tetrahedron shapes, respectively. The $\Lambda$ value is set to zero.

It is known that $^{12}$C is a mixture of shell 
and cluster configurations~\cite{Kanadaenyo07,Chernykh07,Otsuka22}.
To incorporate this mixture, we introduce the mixed configuration (M-type):
we take $\Lambda=0.2$ and $d=2$ fm following the optimal parameters based on
a variational calculation~\cite{AQCM14} and fix the $\nu$ value
to reproduce the charge radius of $^{12}$C.
  For $^{16}$O, here we only investigate
  a simple four-$\alpha$ cluster state with $\Lambda=0$
  because the breaking effect of $\alpha$ clusters
  is rather limited as shown in a $^{12}{\rm C}+\alpha$ model
  calculation~\cite{AQCM07}: The ground state of $^{16}$O is insensitive
  to the strength of the spin-orbit interaction
  because the additional four nucleons compensate the attractive
  effect of the spin-orbit interaction in $^{12}$C.

Table~\ref{results.tab} lists the calculated profiles of these obtained
trial states for $^{12}$C and $^{16}$O:
The fixed parameter sets, the total harmonic oscillator quanta $\left<Q\right>$,
the expectation values of single-particle spin-orbit operators
$\sum_{i=1}^A\bm{l}_i\cdot\bm{s}_i$, $\left<LS\right>$,
single-particle parity operators
$\sum_{i=1}^AP_{i}$ with $P_if(\bm{r}_i)=f(-\bm{r}_i)$,
$\left<P\right>$.
We see that the S-type states nicely reproduce the expected values of
the ideal shell-model configurations, which are given in parentheses.
In the C-type states, the cluster distance in $^{12}$C
is significant and larger than the radius of $\alpha$ particle, 1.46 fm.
The distance becomes larger for $^{16}$O,
showing more developed $\alpha$-cluster structure.

\begin{figure}[ht]
\begin{center}
  \epsfig{file=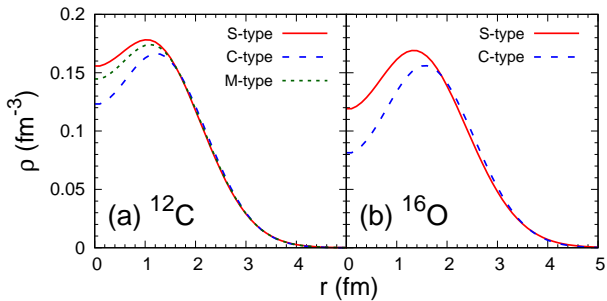, scale=0.85}
  \caption{One-body nucleon density distributions of (a) $^{12}$C and (b) $^{16}$O.}
    \label{dens.fig}
  \end{center}
\end{figure}

This characteristic is well reflected in the density profiles.
Figure~\ref{dens.fig} plots those obtained one-body
density distributions for $^{12}$C and $^{16}$O.
The S-type and C-type exhibit quite different density profiles
despite the fact that  their root-mean-square radii are the same.
In $^{12}$C, the C-type produces more depressed density distributions in the internal regions.
This ``bubble'' structure comes from a well-developed $\alpha$ cluster configuration.
Furthermore, this difference becomes more apparent in $^{16}$O,
reflecting the larger $\alpha$ cluster distance than the C-type of $^{12}$C.
In the surface regions, the C-type density distributions drop more rapidly than the S-type ones.
Since the $\alpha$ particle has an extremely sharp nuclear surface,
the C-type has a sharper surface than the S-type,
reflecting the well-developed $\alpha$ cluster structure,
i.e., the $\alpha$-$\alpha$ inter-cluster distance is large enough.
As expected, the M-type gives the density distribution
intermediate between the S-type and the C-type, as well as its properties,
as shown in Table~\ref{results.tab}.

To evaluate the density profiles near the nuclear surface more quantitatively,
we also calculate the diffuseness parameter
using the prescription given in Ref.~\cite{Hatakeyama18}.
By assuming a two-parameter Fermi (2pF) function as the one-body density distribution
\begin{align}
\rho_{\rm 2pF}(\bar{R},a,r)=\frac{\rho_{0}}{1+\exp[(r-\bar{R})/a]},
\end{align}
the radius $\bar{R}$ and diffuseness $a$ parameters
are determined by minimizing the value
\begin{align}
  \int_{0}^\infty dr\,r^2\left|\rho_{\rm 2pF}(\bar{R},a,r)-\rho(r)\right|.
\end{align}
$\rho_0$ is uniquely determined by the normalization condition.
The extracted diffuseness parameters are 
0.432~fm  (0.401~fm) for the S-type (C-type) of $^{12}$C and
0.455~fm (0.405~fm) for the S-type (C-type) of $^{16}$O.
We again confirm the sharper nuclear surfaces for the C-type quantitatively,
and the difference between the S-type and the
C-type are significant,
which can be distinguished by the proton-nucleus
elastic scattering experiment~\cite{Hatakeyama18}.
The diffuseness parameter for the M-type
is found to be 0.423~fm, showing
an intermediate value between
the S-type and the C-type in $^{12}$C.

\begin{figure}[ht]
  \begin{center}
     \epsfig{file=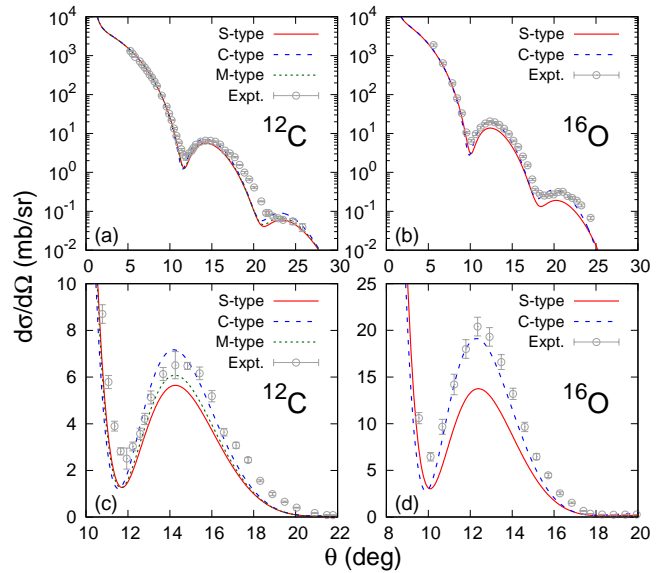, scale=0.85}
  \caption{Proton-nucleus differential elastic scattering cross sections
    for (a) $^{12}{\rm C}$ and (b) $^{16}{\rm O}$
    at incident energies of 1000~MeV as a function of scattering angles.
    The cross sections in a linear scale around the first peak positions
    for (c) $^{12}{\rm C}$ and (d) $^{16}{\rm O}$ are also plotted for visibility. 
    The experimental data is taken from Refs.~\cite{Alkhazov72,Alkhazov85}.}
    \label{dcs.fig}
  \end{center}
\end{figure}

The difference in the surface density profile
can be probed by the proton-nucleus elastic scattering~\cite{Sakaguchi17}.
Differential elastic scattering cross sections at around the first peak position
reflects the surface density profiles and can be a promising probe
to extract information on various nuclear structure~\cite{Hatakeyama18,Choudhary20,Choudhary21,Tanaka20, Horiuchi21b,Horiuchi22,Horiuchi22b}.
Figure~\ref{dcs.fig} plots the proton-nucleus
differential elastic scattering cross sections for $^{12}$C and $^{16}$O.
Incident energy is chosen to be 1000~MeV,
where the experimental data is available both for $^{12}$C and $^{16}$O.
The overall agreement between the theory and experiment is obtained.
To see it more quantitatively, we also plot 
the cross sections around the first peak position in a linear scale
as they are closely related to the density profiles near the nuclear surface.
For $^{12}$C, the S-type underestimates
the cross sections around the first peak,
while the C-type slightly overestimates them but
is closer to the data.
This suggests that the $^{12}$C wave function contains
a significant amount of the $\alpha$-cluster component.
In reality, the $^{12}$C wave function is the mixture
of the shell and $\alpha$-cluster configurations
as the M-type gives better reproduction of the data.
For $^{16}$O, the theoretical cross sections
with the C-type better reproduce the data,
indicating more developed $\alpha$ cluster structure than $^{12}$C.

\begin{figure}[ht]
  \begin{center}
    \epsfig{file=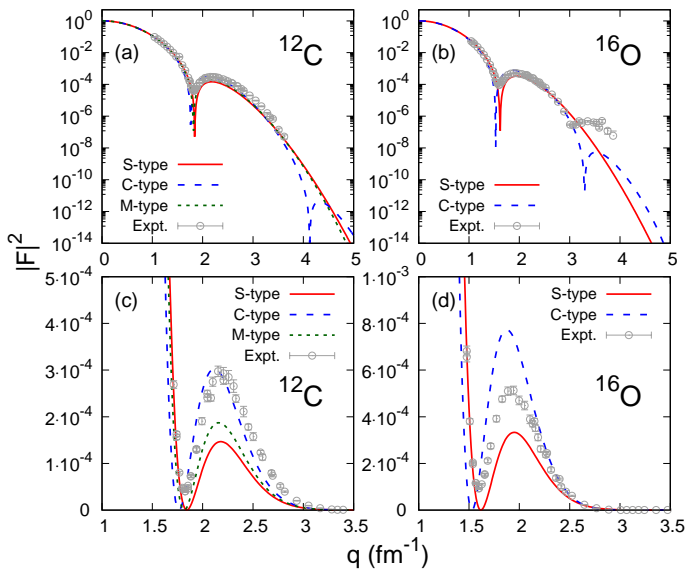, scale=0.85}
  \caption{Squared elastic charge form factors of $^{12}$C and $^{16}$O
    as a function of the momentum transfer.
  The experimental data is taken from Ref.~\cite{Sick70}.}
    \label{form.fig}
  \end{center}
\end{figure}

The one-body density distributions have traditionally been investigated by
electron scattering~\cite{Hofstadter56}.
As it probes different parts of the density distribution,
it is interesting to compare the elastic charge form factors
obtained from the present trial states.
The elastic charge form factor of the $N\alpha$ cluster model
is evaluated by the Fourier transform
of the density distribution with a finite size effect of a proton
charge as~\cite{Kamimura81}
\begin{align}
  |F(q)|^2=\left|\frac{4\pi}{A}\int_0^{\infty}dr\, \rho(r)j_0(qr) r^2\right|^2
  \exp\left(-\frac{1}{2}a_p^2q^2\right)
\end{align}
with $a_p^2=0.514$ fm$^2$, which reproduces
the charge radius of a proton, $0.878$ fm~\cite{Angeli13}.
Figure~\ref{form.fig} displays the square of the elastic charge form factors
for $^{12}$C and $^{16}$O as a function of the momentum transfer $q$
both in logarithmic and linear scales.
As already seen in the differential elastic scattering cross sections,
a comparison with the experimental data supports a mixture
of the shell and $\alpha$-cluster components both in $^{12}$C and $^{16}$O.
The differences in the charge form factor become more evident at higher
momentum transfer $q$.
For $^{16}$O, no constraint is obtained
in the present comparison of the data at $q \gtrsim 3~$fm$^{-1}$.

We confirmed a significant amount of $\alpha$-cluster components
are included in the ground state wave functions in $^{12}$C and $^{16}$O.
As seen in Fig.~\ref{dens.fig}, the characteristics of the density profiles
are not only for the surface region
but also for internal regions. Since the electric force is much weaker than
the nuclear force, the elastic charge form factor probes more internal regions
than the proton-nucleus elastic scattering.
Only the surface regions around the half density can be reflected~\cite{Hatakeyama18,Choudhary20} in the proton-nucleus elastic scattering.
Though the sensitivity on the density profiles is different in
proton or electron probes, we can see distinctive patterns enough
to distinguish the shell and cluster configurations
in the higher momentum transfer regions beyond the first peak.
Such measurements up to the second peak could
be useful for investigating both the surface and internal density profiles.

In summary, to settle the controversy on the dominance of the shell
or cluster configurations,
we have investigated the one-body
density profiles of $^{12}$C and $^{16}$O
using the basis states of the antisymmetrized quasi-cluster model,
which can describe both the shell and $\alpha$-cluster configurations
in a single scheme.
Despite that, the shell- and cluster-model type wave functions reproduce
the same charge radius, these density profiles show
quite different characteristics.
As the $\alpha$ cluster structure is developed,
the density distribution exhibits
sharper nuclear surface and more depressed central densities
compared to the shell-model configuration.

We find, for the first time,
that the evidence of the $\alpha$ clustering
is imprinted on the one-body density profiles
by comparing the proton-nucleus elastic scattering cross sections
and the elastic charge form factors.
The cross sections or charge form factors
near the first peaks are significantly enhanced
when the ground state wave function includes the $\alpha$ cluster component.
The nuclear surface density profiles can be distinguished
by measuring these quantities as a function of the momentum transfer
up to the first peak position.
It should be noted that 
the proton-nucleus scattering has
the advantage that it can be more easily applied
to unstable nuclei using inverse kinematics~\cite{Matsuda13},
though electron scattering
for unstable nuclei is available~\cite{Tsukada17}.
Applications of the present method to heavier $4N$ systems
as well as neutron-rich unstable nuclei are
may also help for a more universal understanding of
the competition between the shell and cluster configurations.

This work was in part supported by JSPS KAKENHI Grants
Nos.\ 18K03635, 22H01214, and 22K03618.
We acknowledge the Collaborative Research Program 2022, 
Information Initiative Center, Hokkaido University.

\end{document}